\documentclass[twocolumn]{aastex62}
\usepackage{latexsym}
\usepackage{color}
\usepackage{amsmath}
\usepackage{amssymb}
\usepackage{graphicx}
\usepackage{aas_macros}

\newcommand{\Ms} {$\rm{M_{\odot}}~$}

\usepackage{hyperref}

\submitjournal{ApJ}
\shortauthors{Schauer et al.}
\shorttitle{The ULT - requirements for observing Pop III stars}

\begin{document}

\title[The ULT - requirements for observing Pop III stars]
{The Ultimately Large Telescope - what kind of facility do we need to detect Population III stars? } 

\correspondingauthor{Anna T. P. Schauer}
\email{anna.schauer@utexas.edu}

\author[0000-0002-2220-8086]{Anna T. P. Schauer}
\altaffiliation{Hubble Fellow}
\affiliation{Department of Astronomy, 
University of Texas at Austin,
TX 78712, USA}

\author[0000-0002-7339-3170]{Niv Drory}
\affiliation{McDonald Observatory, 
University of Texas at Austin,
TX 78712, USA}

\author{Volker Bromm}
\affiliation{Department of Astronomy,
University of Texas at Austin,
TX 78712, USA} 

\begin{abstract}
The launch of the James Webb Space Telescope will open up a new window for observations at the highest redshifts, reaching out to $z\approx 15$. However, even with this new facility, the first stars will remain out of reach, as they are born in small minihalos with luminosities too faint to be detected even by the longest exposure times. 
In this Letter, we investigate the basic properties of the \textit{Ultimately Large Telescope}, a facility that can detect Population III star formation regions at high redshift. Observations will take place in the near-infrared and therefore a moon-based facility is proposed. 
An instrument needs to reach magnitudes as faint as 39\,mag$_\mathrm{AB}$, corresponding to a primary mirror size of about 100\,m in diameter. Assuming JWST NIRCam filters, we estimate that Pop~III sources will have unique signatures in a colour-colour space and can be identified unambiguously. 
\end{abstract}

. 
\section{Introduction}\label{introduction}
The current record-holding observation of a high redshift galaxy was performed by 
\cite{oesch16}, reaching $z \sim 11{}$ with a stellar mass of 
$M_\star \sim 2\times  10^9{}$\,\Ms and a star formation rate of 
$\sim 25$\,\Ms\,yr$^{-1}$. 
The upcoming James Webb Space Telescope (JWST) is targeting the first galaxies and expected to push this barrier to even earlier times. The JWST targets high-redshift sources, as their emitted light is redshifted into the near-infrared (NIR) wavelength bands. JWST filters reach flux limits of a few nJy, thus opening up the observational window for galaxies out to a redshift of $z\ge15$ \citep{gardner06}. Small galaxies with stellar masses of a few $\ge10^7$\,\Ms will be detectable out to redshifts of $z\sim 12$ \citep{barrow17,ceverino19} in relatively large numbers (e.g.\ 60 objects in a single deep pointing between $z=8-11$; \citealt{hainline20}). 

However, to understand the formation of the first stars and the early Universe as a whole, we need to observe so-called minihalos, the formation site of Population~III (Pop~III) stars with virial masses of $10^5 - 10^7$\,\Ms \citep{hir15,anna19}. The luminosities of these objects are much smaller, and we expect a total mass in Pop~III stars of a few hundred to a few thousand solar masses \citep{xu13,hir14}. 
None of the existing nor planned missions will be able to detect an individual minihalo with a standard Pop~III stellar mass \citep{zack11,zack15}, even more massive first galaxies at $z\sim8$ with $M_\star < 10^6$\,\Ms are out of reach for JWST \citep{jeon19}, unless observed during a starburst. 

In this Letter, we investigate what kind of telescope is necessary to 
observe such minihalos, and therefore the formation sites of Pop~III stars or star clusters. 
Two physical parameters are important: the number density of such minihalos, translating into constraints on the 
field of view, and the stellar properties that determine the luminosity of a single object, and our analysis will focus on them. With such a next-generation, {\it Ultimately Large Telescope} (ULT), the entire cosmic star formation history would become accessible, completing astronomy's centuries old pursuit to push the frontier of the observable universe.  

\section{Minihalo models}\label{methods}
\subsection{Stellar Populations}
The mass of the first stars is still under debate. 
Earlier pioneering studies showed that a single, very massive Pop III star forms per minihalo with masses reaching 300-1000\,\Ms \citep{abn02,bcl02}. 
In more recent simulations, however, the central gas core exhibits a 
rotating disk that is breaking up into several lower-mass stars (\citealt{get11,clark11,stacy16}; see however \citealt{hir14}). 
The resulting spectral energy distribution (SED) of the stars depends on additional input physics that is uncertain, such as the contribution from nebula emission \citep{zack11}. 
To account for the different possibilities, we study a range of models, including two limiting cases for Pop~III: 
\begin{itemize}
    \item \textit{Pop~III MS star}: We assume that a single, massive star forms with a mass of $\sim$1000\,M$_{\odot}$, staying on the main sequence (MS) throughout. As the luminosity per stellar mass is approximately constant for stars $>$100\,\Ms \citep{bromm01}, it is unimportant if we have, say, three 300\,\Ms stars or one 900\,\Ms star. We use the spectrum in \cite{bromm01}, which does not take into account nebula emission. 
    \item \textit{Pop III + nebula + evolution}: To mimic a stellar population that is composed of very massive stars, we choose the Pop~III.1 Yggdrasil model by \cite{zack11} with an extremely top-heavy initial mass function (IMF), within 50-500\,M$_\odot$ and a Salpeter slope, including nebula emission, and the effect of stellar evolution. 
    \item \textit{Pop~II SED}: For a Pop~II counterpart, we assume that a small cluster of stars forms with continuous star formation over 30\,Myr in each minihalo. Here, we choose the Pop~II Yggdrasil model \citep{zack11} with a Kroupa IMF between 0.1 -- 100\,M$_\odot$, a covering fraction of 50\% (equivalent to a nebula emission fraction of 50\%) and a metalicity of $Z = 0.0004Z_\odot$ (corresponding roughly to the critical metallicity $Z=10^{-3.5}\,Z_\odot{}$ that divides the Pop~III from the Pop~II formation regime, see \citealt{bromm01b}). 
\end{itemize}

As the total mass in stars per minihalo varies, we have some fundamental scatter (which, in nature, also comes from different realisations of the IMF). 
For this proof-of-concept paper, we take a simplistic approach, where we set the fiducial mass per minihalo to 1000\,M$_\odot$, and assume that the stellar mass follows a logarithmically normal distribution with a standard deviation of 0.5\,dex. 
As typical star forming minihalos have a virial mass of around $10^6$\,\Ms \citep{anna19b}, our fiducial star formation efficiency (SFE) is 0.1\%, with 95\% of halos having a SFE between 0.01\% (accounting for the single star case in a minihalo) and 1\% (accounting for efficient SF). 
These numbers lie in the range of SF limits inferred from the 21cm-signal \citep{fialkov19,anna19}. 

The \textit{Pop~III MS star} model is time-independent. However, the two other models change their flux over time, as more massive stars have a shorter lifetime and the nebula reprocessing changes. Below, we show the flux either explicitly at defined times, or we consider average fluxes over the entire time evolution, e.g. when discussing the fraction of observable sources. 
%
\subsection{Number of luminous minihalos}
\begin{figure}
    \centering
    \includegraphics[width=0.99\columnwidth]{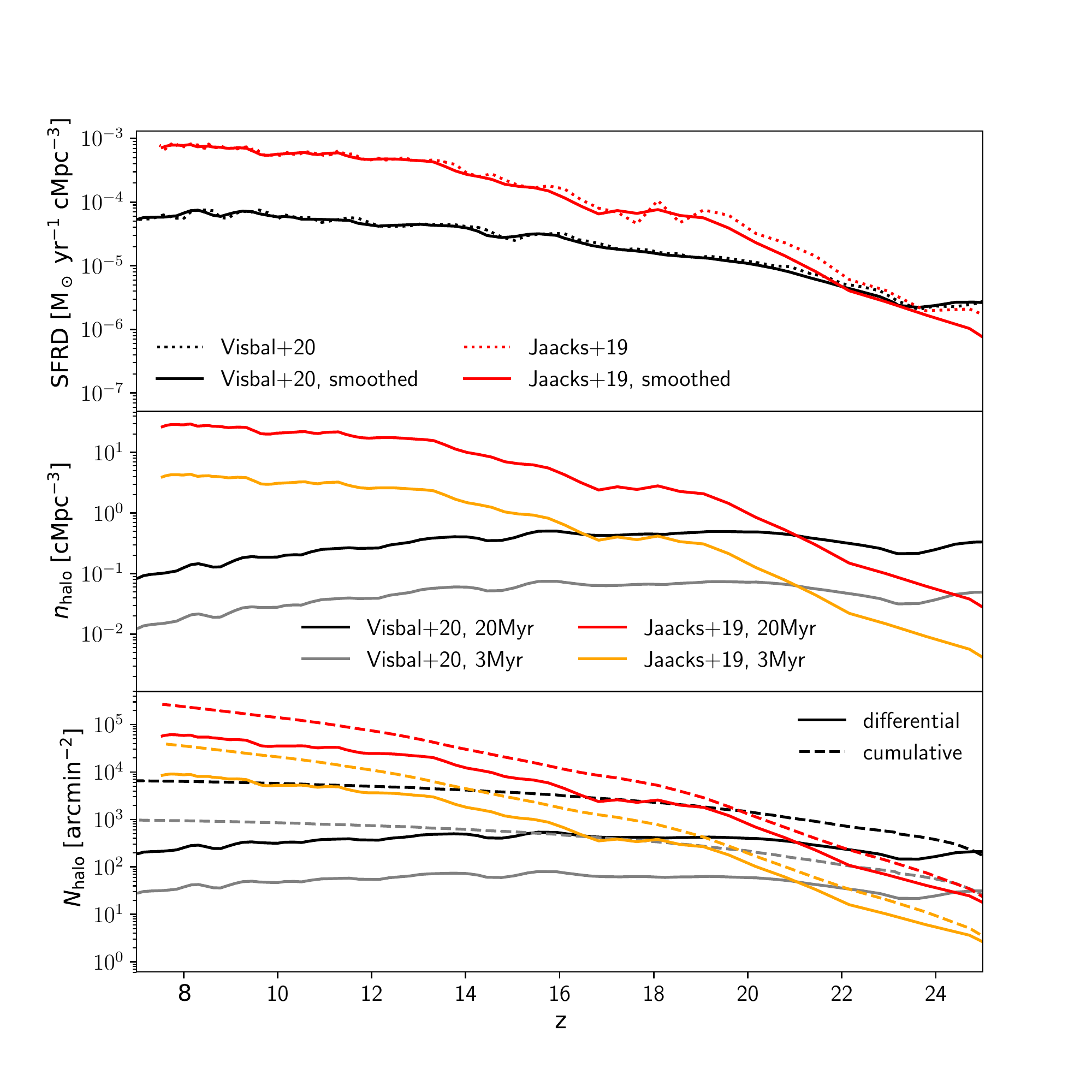}
    \caption{Top panel: Pop~III SFRD from select simulations (J19 and V20, see main text). We show both the original data (dotted lines) and after we applied a smoothing algorithm (solid lines). Middle panel: Number density of luminous halos per comoving volume, derived from the SFRDs. Bottom panel: Number of luminous halos per observed field of view, normalized to 1~arcmin$^2$.}
    \label{fig:fov}
\end{figure}

For the number density of minihalos that currently host Pop~III stars, 
we have to rely on theoretical models. 
We take some of the most recent simulation results and work ``backwards'' from 
a star formation rate density (SFRD) to obtain the number of luminous minihalos per redshift.
We choose \cite{jaacks19} (J19) and \cite{visbal20} (V20), who present global Pop~III SFRDs based on their simulations (see top panel of Fig.~\ref{fig:fov}). In their models, they assume a transition to Pop II at critical metallicities of $10^{-4}\,Z_\odot$ and $10^{-3.5}\,Z_\odot$, values often assumed in the literature.

We can translate the SFRD, $\Psi(z){}$, into 
a comoving number density of halos with luminous Pop~III stars via
\begin{equation}
   n_\mathrm{halo}(z) = \Psi(z) \times t_\star / M_\star(z)\mbox{\ ,}
\end{equation}
where $t_\star$ is the stellar lifetime. We consider two values for $t_\star$ \citep{schae02}:  20\,Myr (lifetime of a 9\Ms star) and 3\,Myr (lifetime of a 80\Ms star). For $M_\star(z)$ in J19, we take their 
fiducial average value of $M_\star^\mathrm{ave}\sim550$\,M$_{\odot}$. 
V20 links the stellar mass to the halo mass, via $M_\star (z) = 0.001 (\Omega_b/\Omega_m) \times M_\mathrm{vir}(z)$. We retrace their steps for the minimum halo mass, and choose the most likely streaming velocity region of the universe with $v_\mathrm{bc} = 0.8\sigma$ \citep[see e.g.][for a review on streaming velocities]{fialkovreview14}. 

The resulting halo mass functions of luminous halos are shown in the middle panel of Figure \ref{fig:fov}. We see a large spread in minihalo number density, which is mainly due to the two different star formation 
descriptions. While $M_\star{}$ is constant in J19, its mass increases 
by more than one order of magnitude in V20 for lower redshifts. The two models hence span a wide range of physical possibilities. 

\section{Observing Pop~III stars}\label{obs}

\subsection{Field of view}\label{fov}
We can translate the number density of luminous Pop~III halos, $n_\mathrm{halo}{}$, into 
the expected number of luminous halos per arcmin$^2$ on the sky, $N_\mathrm{halo}$. Integrating over the redshift interval $z_1{}$ to $z_2{}$, we have: 

\begin{equation}
N_\mathrm{halo} = \int_{z_\mathrm{min}}^{z_\mathrm{max}} \mathrm{d}z\, 
n_\mathrm{halo} \frac{ c (1+z)^2 d_A(z)^2} {H_0 E(z)} \Delta \Omega \mbox{\ .}
\end{equation}{}
We show our results for the J19 and V20 data in the bottom panel of Figure \ref{fig:fov}. Even for the 
short lifetime of the 80\,\Ms star within the V20 model, we expect to observe more than 20 minihalos per square arcminute per unit redshift. 
Comparing, e.g., to the $\sim 10$ square arcminute NIRCam field of view on JWST, 
we can conclude that there will be a large number of objects visible, and that the detection of minihalos will never be limited by too low a number of sources, as long as the ULT is able to probe sources that are faint enough. 

\subsection{Sensitivity}\label{sens}
As a next step, we calculate the observed flux of our Pop~III stellar sources. 
As seen in more recent simulations \citep{chiaki18}, the ionization front around the central source hardly breaks out of the virial radius, and therefore 
the nebula reprocessing assumed in the \textit{Pop~III + nebula + evolution} model is the more realistic case. 
The upcoming JWST is designed to probe sources at high redshifts with the NIRCam instrument \citep{rieke05,beichhman12}. For simplicity, we adopt 
the same filters and assume the same filter throughput as in the NIRCam wide-field 
filters. We assume that the filter technology will only advance towards higher sensitivities, such that our results can be seen as conservative limits. 

From the Yggdrasil-luminosities $L_\lambda$, we can infer the observed flux of the star or stellar population: 
\begin{equation}
f_\lambda = \frac{L_\lambda}{4 \pi d_c^2\, (1+z)^3} \mbox{\ ,} 
\end{equation}
where $d_c{}$ is the comoving distance to redshift $z$. Applying the 
NIRCam-filter response functions, $R(\lambda)$, the integrated 
flux in the observed frame is: 
\begin{equation}
<F_\lambda> = \frac{\int \lambda f_\lambda R(\lambda) \mathrm{d}\lambda}{\int \lambda  R(\lambda) \mathrm{d}\lambda} \mbox{\ .}
\end{equation}

As most sensitivity limits refer to specific frequency, we can 
translate our observed flux $<F_\lambda>{}$ from wavelength to 
frequency space $<F_\nu>{}$ via the pivot wavelength: 
\begin{equation}
    <F_\nu> = 3.34\cdot10^{10} \mathrm{Jy} \left(\frac{\lambda}{\mu\mathrm{m}} \right)^2 <F_\lambda> .
\end{equation}

\begin{figure}
    \centering
    \includegraphics[width=0.99\columnwidth]{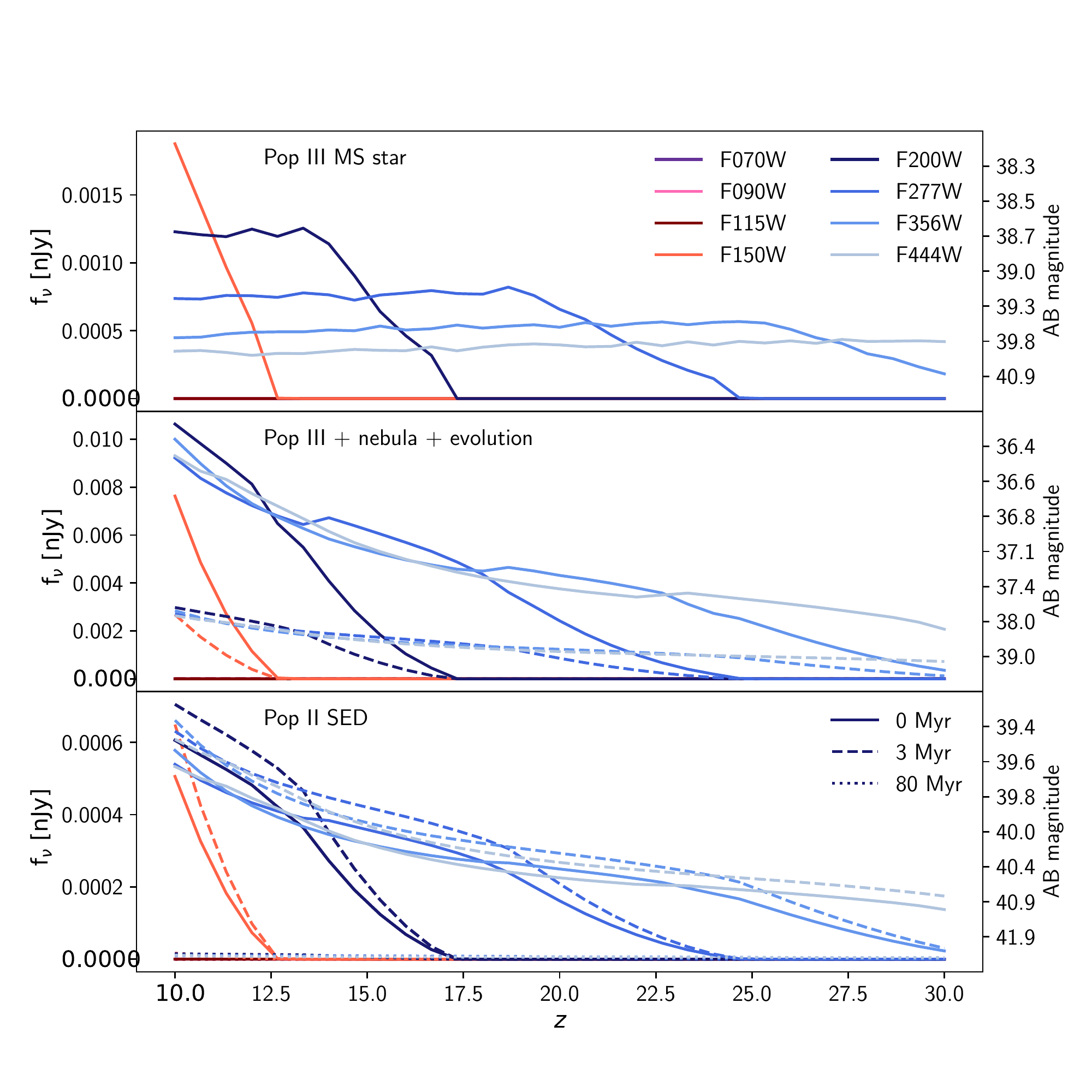}
    \caption{Observed flux as a function of redshift for our three stellar population models. For the \textit{Pop~III + nebula + evolution} and the \textit{Pop~II SED} models, we include two (or three) time intervals: 0 Myr, 3Myr (and 80Myr).}
    \label{fig:flux}
\end{figure}

We show the flux in the NIRCam filters as a function of redshift in Figure \ref{fig:flux} for our three stellar models.
For higher redshifts, the flux in the longer wavelength filters 
becomes the dominant component. To enable observations at redshift 
$z \ge 15$, one needs a filter with a wavelength of at least 
$\lambda \ge 2.0\,\mu$m. 
The three different models vary by more than an order of 
magnitude in expected flux. The Pop~III SED with a nebular contribution (middle panel of Fig.~\ref{fig:flux}) has the highest flux. The stars without nebula (top panel of Fig.~\ref{fig:flux}) are less luminous at wavelength larger than Lyman-$\alpha$, and the Pop~II model that includes many less-luminous stars (bottom panel of Fig.~\ref{fig:flux}) is the faintest. 
As shown in Section \ref{feasibility}, a telescope with a detection limit of 39 mag$_\mathrm{AB}{}$ can be feasible. Here again, we use the NIRCam filter specifications for our estimates. For example, the F150 filter has a 0.17 lower magnitude limit than the F200 filter, so in our case, m$_\mathrm{AB}^{200} = 39\, \mathrm{mag}{}$ leads to m$_\mathrm{AB}^{150} = 38.83\, \mathrm{mag}{}$. The other filter we will base our predictions on is F444, with m$_\mathrm{AB}^{444} = 37.965\, \mathrm{mag}$. 

\begin{figure*}
    \centering
    \includegraphics[width=0.66\columnwidth]{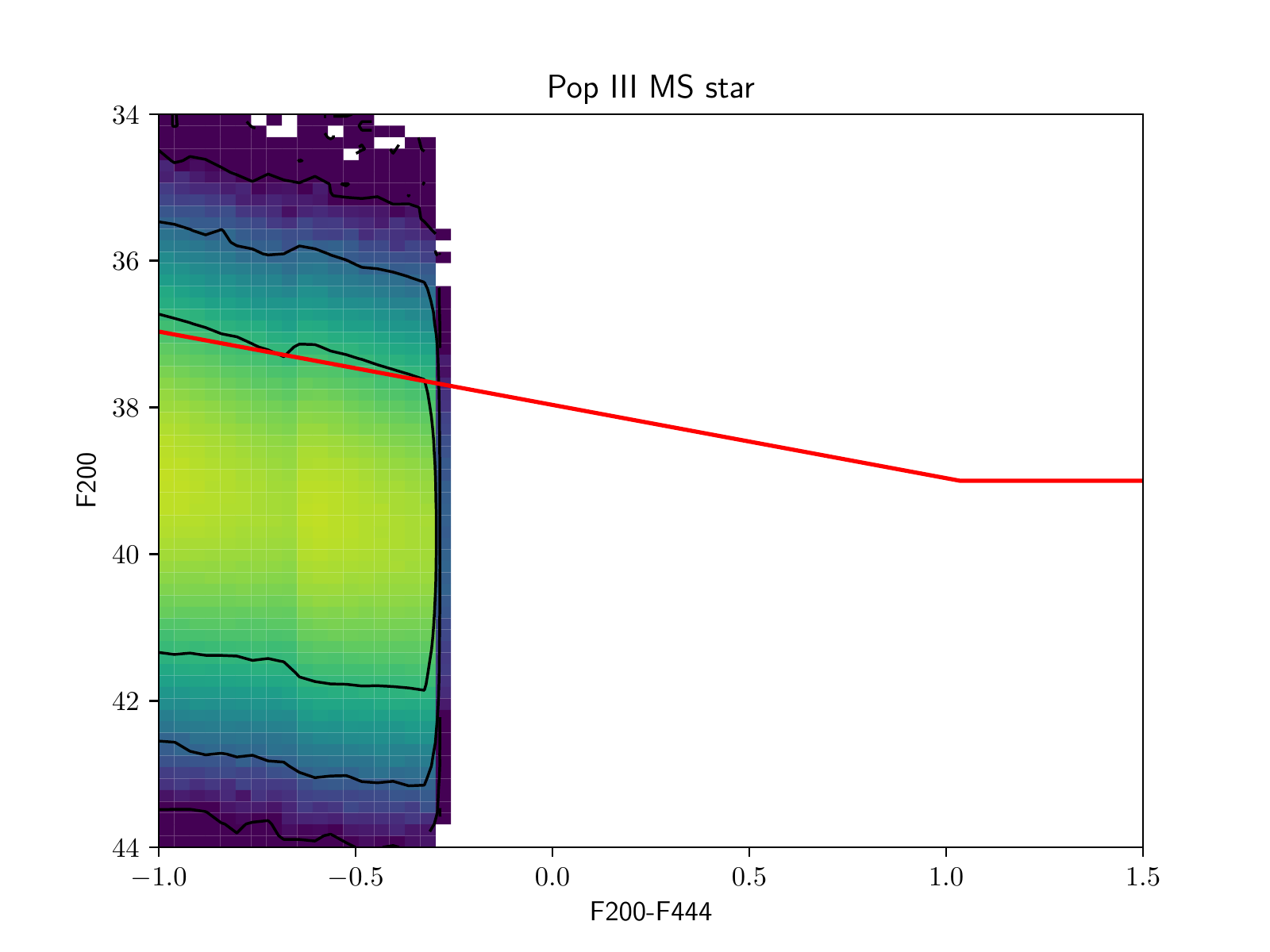}
    \includegraphics[width=0.66\columnwidth]{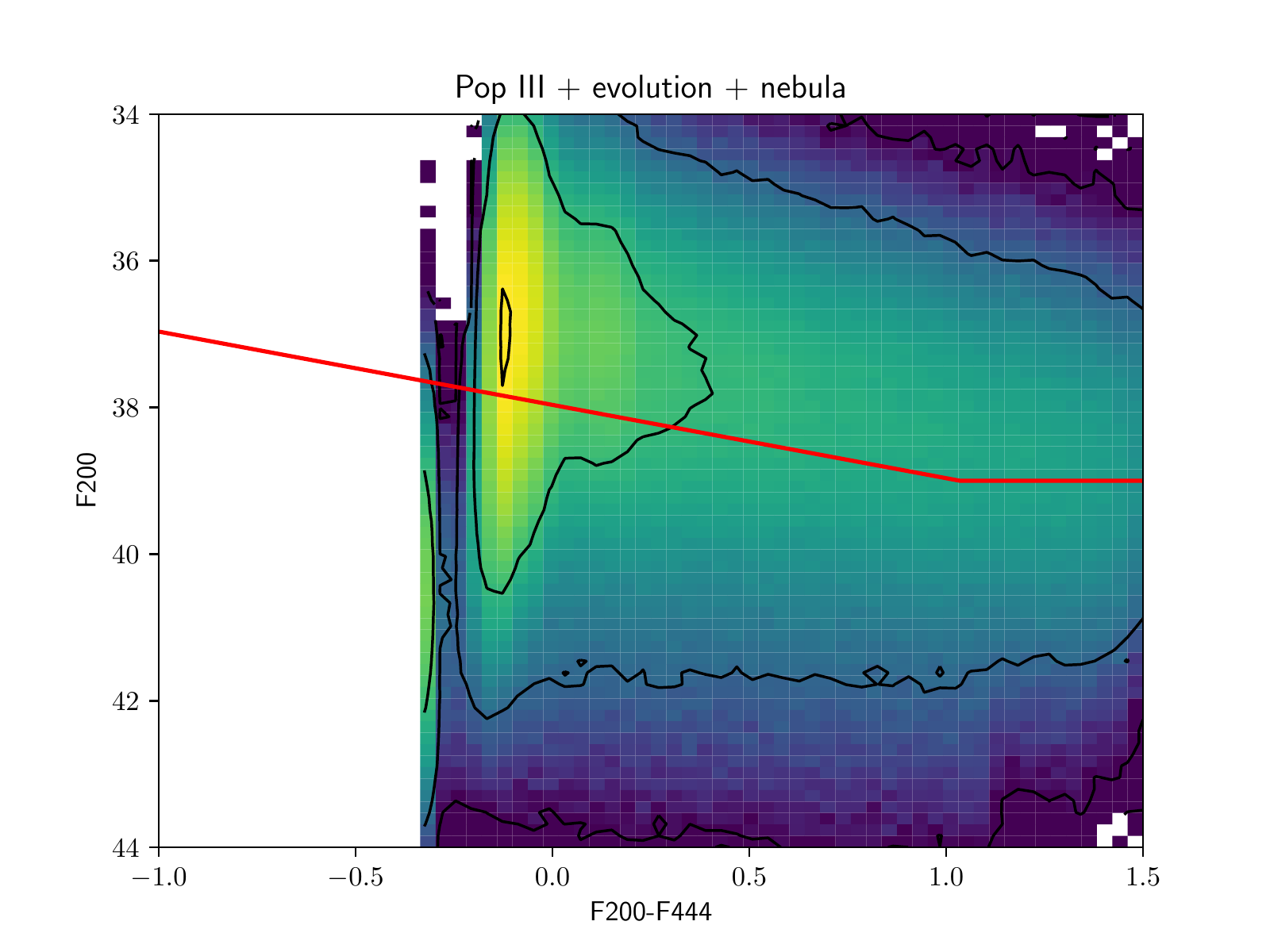}
    \includegraphics[width=0.66\columnwidth]{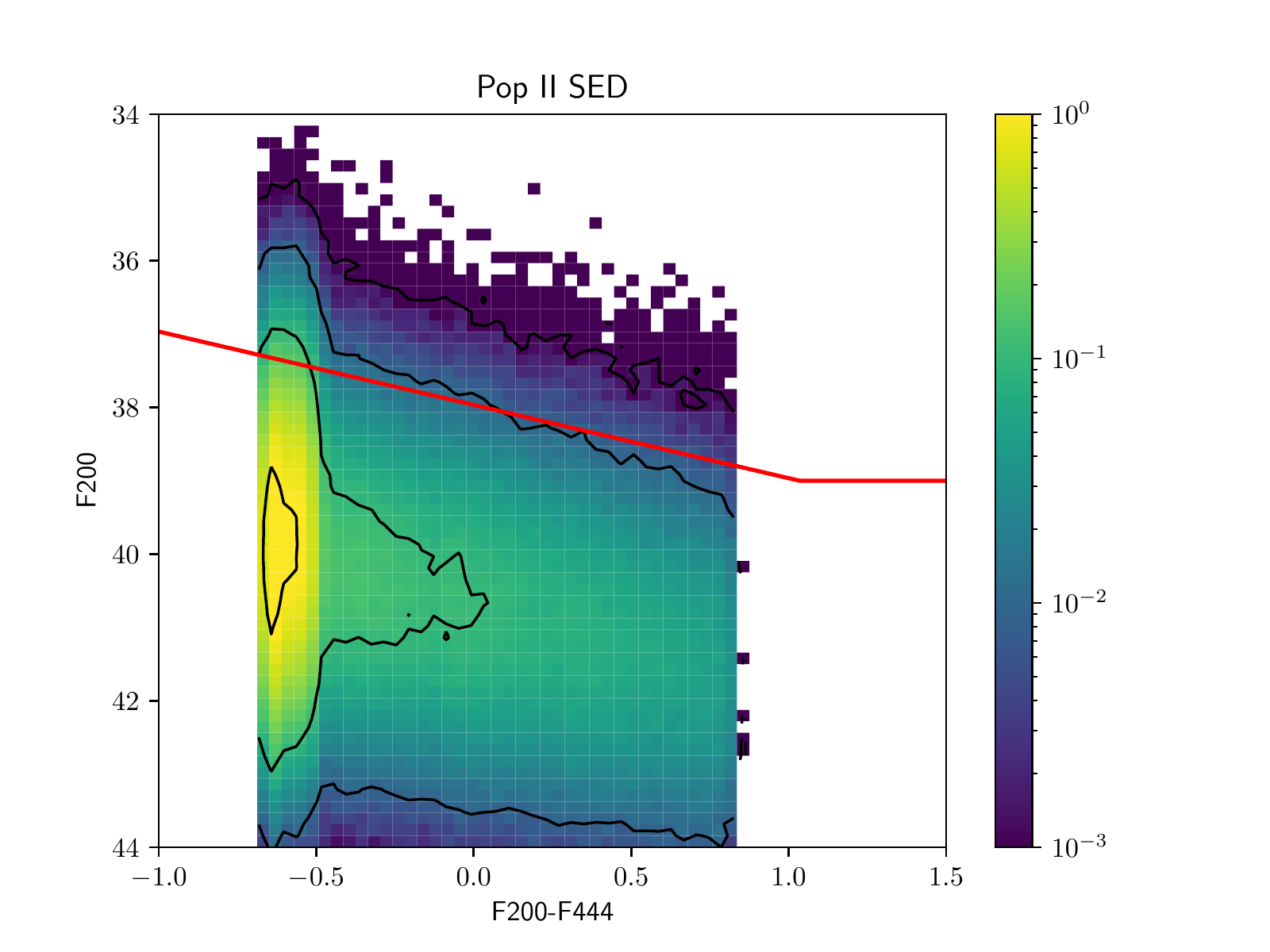}
    \includegraphics[width=0.66\columnwidth]{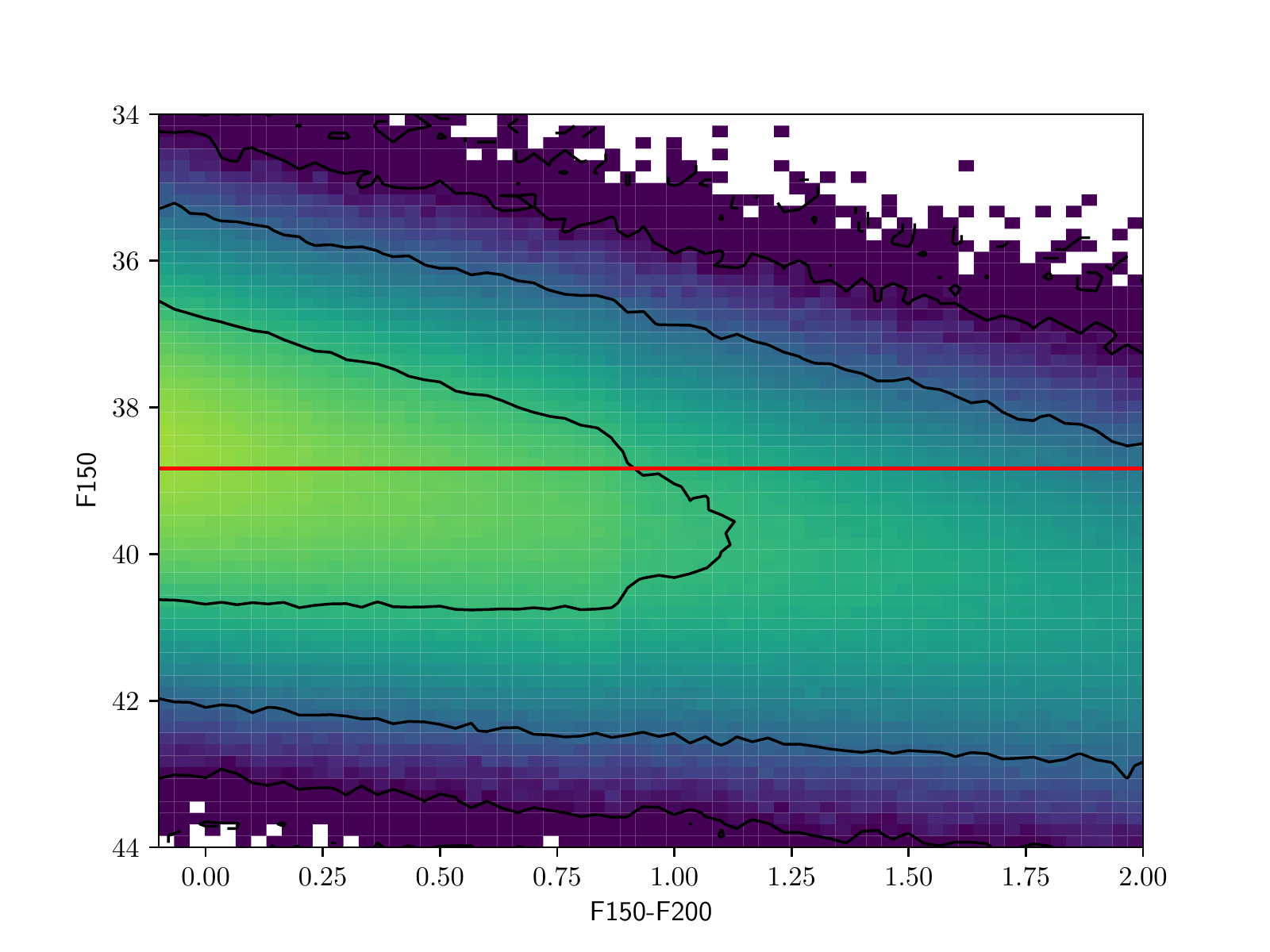}
    \includegraphics[width=0.66\columnwidth]{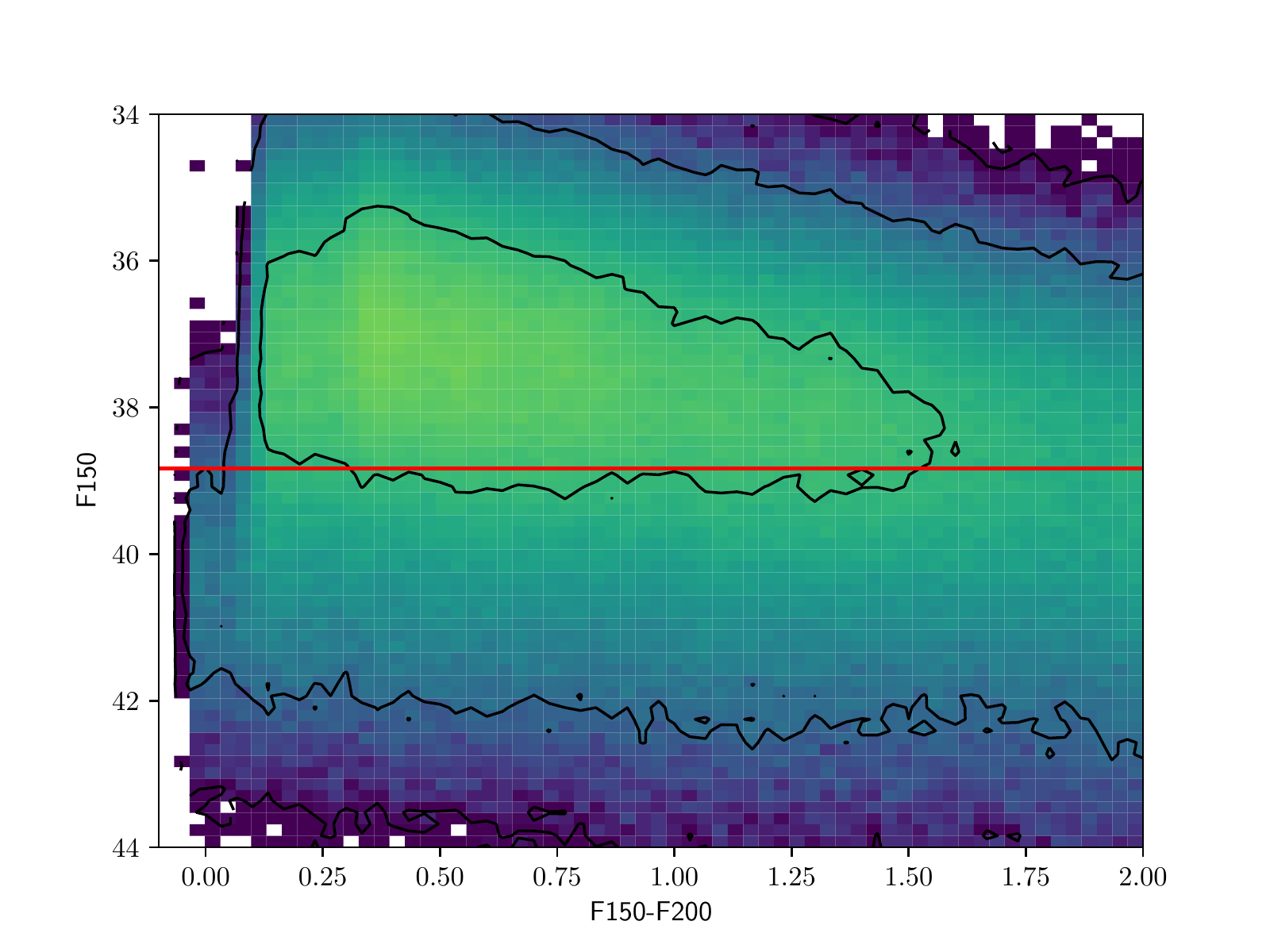}
    \includegraphics[width=0.66\columnwidth]{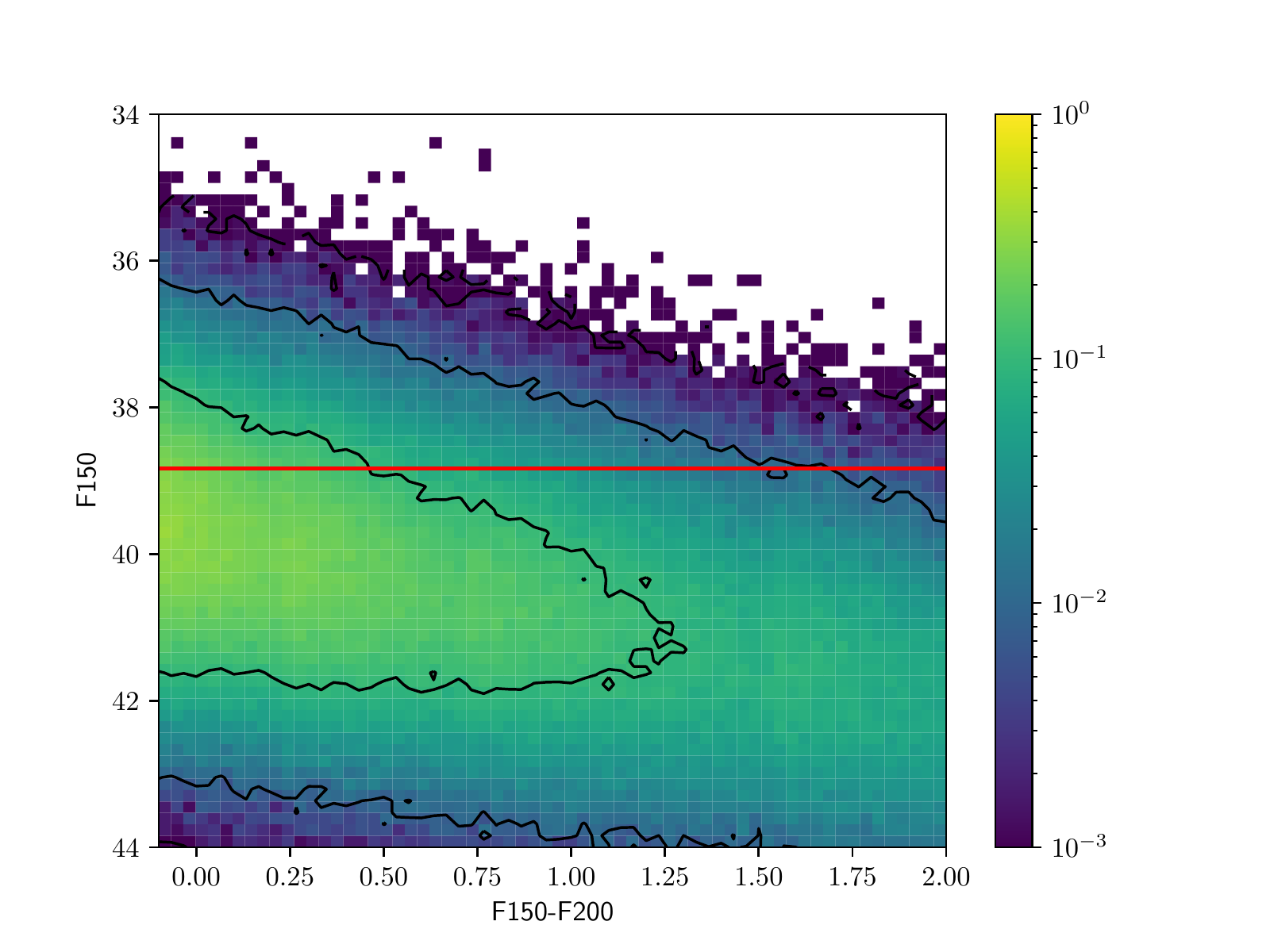}
    \caption{Colour-magnitude density diagrams for our three models. The top row shows the longer wavelength filter combination of F200 and F444 , the bottom row the shorter wavelength filter combination of F150 and F200. Here, we include the redshift interval $z=10$ to $13$, as higher redshifts drop out of the F150 filter. Lower redshifts can be found towards the left of all panels, and higher redshifts towards the right. All sources above the solid red line would be observable with our proposed ULT. }
    \label{fig:cmag}
\end{figure*}

In Figure~\ref{fig:cmag}, we show the density distribution of our three stellar models in colour-magnitude diagrams for the neighbouring filter pair F200 and F444 (top row) and for the distant combination of F150 and F200 (bottom row). 
One can immediately see that for both filter combinations, the majority of sources in the two Pop~III models can be observed (lie above the red solid line), whereas  very few Pop~II sources are bright enough in both bands. 
As the drop-out of the F150-filter occurs around $z\approx13$ (see Fig.~\ref{fig:flux}), we show our sources in the redshift range $10 \leqslant z \leqslant 13 $

\begin{figure*}
    \centering
    \includegraphics[width=1.99\columnwidth]{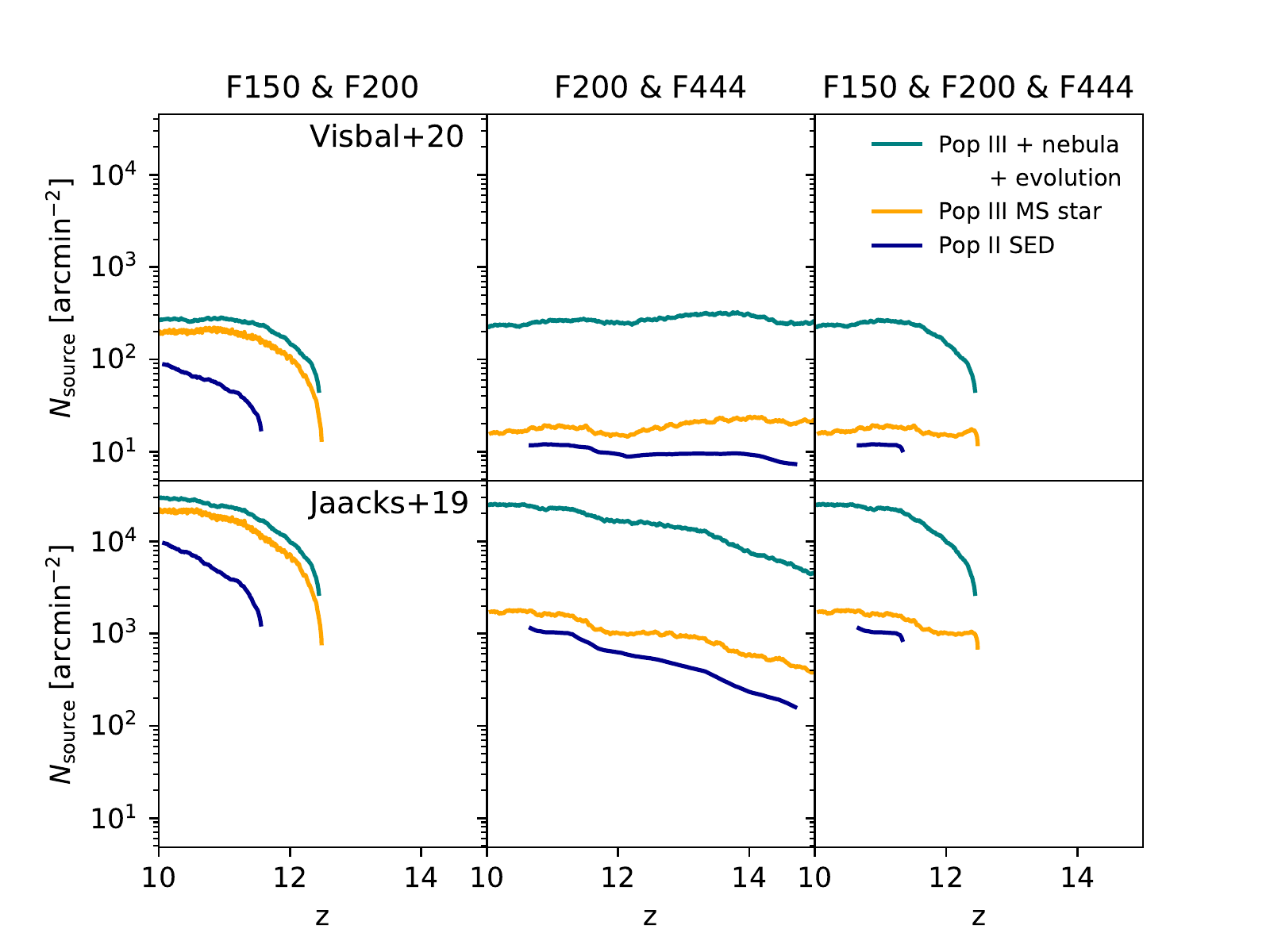}
    \caption{Number density of luminous sources for the two SFRD models by  \cite{visbal20} (top panels) and \cite{jaacks19} (bottom panels). On the left, we show the F150-F200 filter combination, in the middle, the F200-F444 filter combination and on the right the combination of all three filters.}
    \label{fig:lumnum}
\end{figure*}

The estimated number of observable objects with ULT as a function of redshift can be seen in Figure \ref{fig:lumnum}. For the lower wavelength filters (left panels), the drop-out happens around redshift $z\approx12.5$, whereas the number remains roughly constant for the longer wavelength filters (middle panels). Both filter pairs combined (right panels) reflect the drop-out at $z\approx 12.5$, and Pop~II sources are only visible for a short redshift interval around $z=11$. As noted before, the number of detectable Pop~II sources is much smaller than for either the \textit{Pop~III MS star} or \textit{Pop~III + nebula + evolution} model.  

\begin{figure*}
    \centering
    \includegraphics[width=1.99\columnwidth]{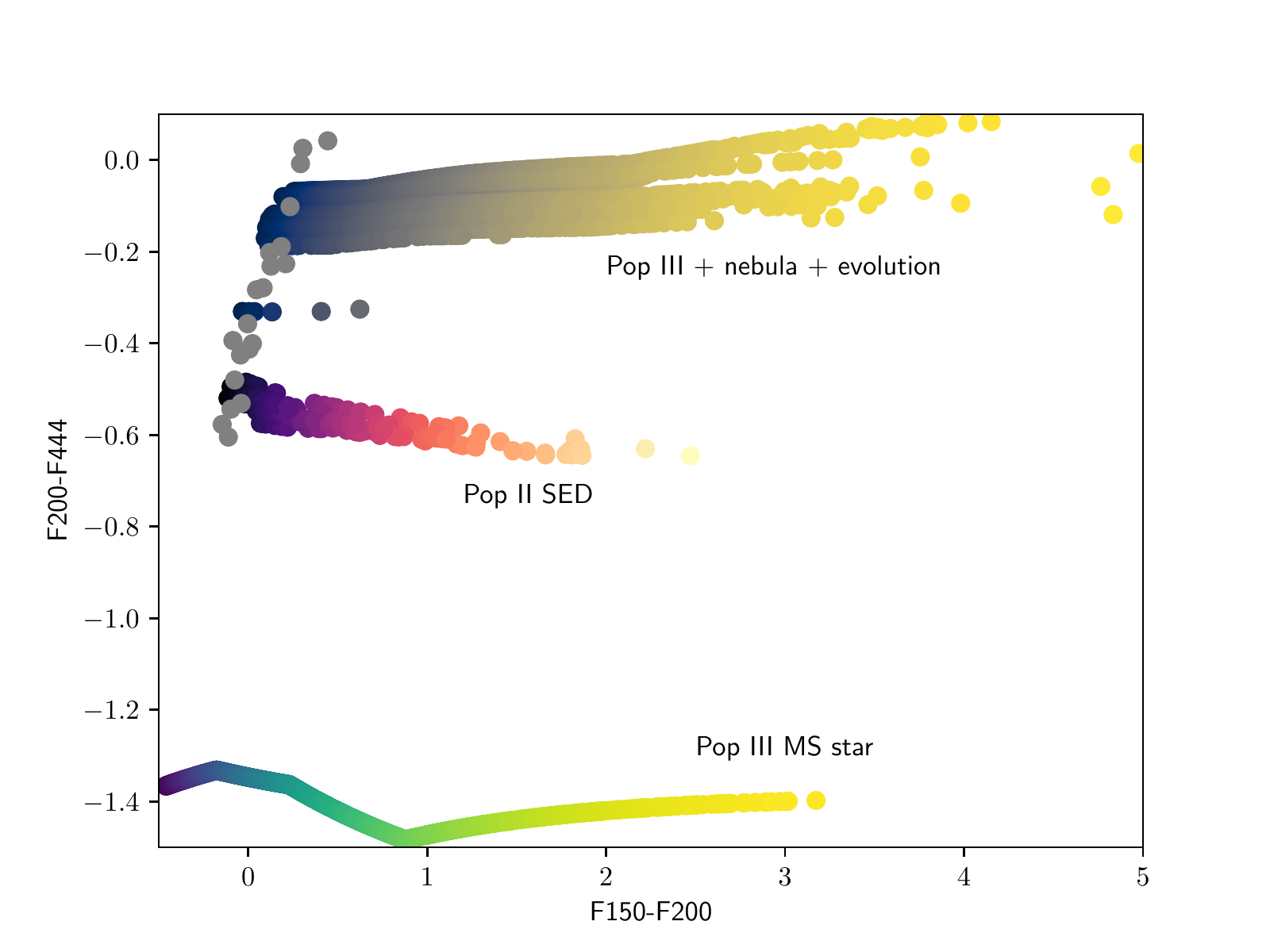}
    \caption{Colour-colour diagram of our three models, from $z=10$ (blue) to $z=13$ (yellow). The \textit{Pop~III + nebula + evolution} and the \textit{Pop~II SED} models consider stellar evolution and are therefore more spread out than the \textit{Pop~III MS star} model. All three high-redshift models occupy distinct regions in this colour-colour diagram and can be distinguished. Overplotted in grey are very cloudy brown dwarfs from \citet{saumon08}  - other brown dwarfs have a positive F200-F400 colour and lie outside the range considered here. }
    \label{fig:cc}
\end{figure*}

Finally, for further diagnostic power, we show a colour-colour-diagram in Figure~\ref{fig:cc}. 
Our three models occupy distinct regions in this diagram, composed of the three filters F150, F200 and F444. 
The different positions of the two Pop~III models are mainly the result of the nebula emission that is present in the \textit{Pop~III + nebula + evolution} model, but not in the \textit{Pop~III MS star} one. With our two Pop~III models, we span the whole parameter space between maximal and zero nebula contribution. Both of these extreme cases can be detected with the ULT.  
With the proposed ULT, we can therefore distinguish between different Pop~III host environments. The \textit{Pop~II SED} model lies between these extremes, and may thus masquerade as a Pop~III model with partial nebula reprocessing. However, as shown in Figure~\ref{fig:cmag}, the Pop~II sources have much lower flux, especially in the F200 filter. 

\subsection{Confusion from other sources}\label{confusion}
In addition to deriving the ULT capabilities for observing Pop~III stellar regions
in minihalos, 
we need to make sure that the sources are unique and cannot be confused with other celestial objects. 
We evaluate a list of possible contaminants with literature data: 
\begin{itemize}
    \item \textit{Pop~II galaxies:} Pop~II galaxies are the first galaxies forming out of metal-enriched gas, and are typically more massive than minihalos. To first order, we can distinguish Pop~II galaxies by dint of their typically lower redshift with the drop-out technique. E.g., $z\ge 9.8$\,galaxies would drop out of the F115W filter \citep{hainline20}. 
    For Pop~II stars forming in galaxies 
    at similar redshift as the considered Pop~III models,
    we can use the colour-colour diagram of the 
    \textit{Pop~II SED} and compare it to the 
    \textit{Pop~III MS star} and \textit{Pop~III + nebula + evolution} models. The Pop~II sources occupy a distinct region in colour-colour space, and are in addition much fainter than the two Pop~III models. 
    \item \textit{Pair-instability supernovae (PISNe):} Some of the brightest sources in the universe are single PISNe, with bolometric luminosities outshining an underlying Pop~III stellar population by a factor of $\sim$\,100. The brightness of these sources declines rapidly, however, and $\sim$\,1 year after the explosion, they are indistinguishable from the underlying stellar population. Even with an optimal strategy, 50,000 exposures per year of 600\,s each are necessary to detect one PISN with JWST \citep{hartwig18}. Given that we target a small field of view, confusion with a PISN is highly unlikely.  
    \item \textit{Brown Dwarfs:} Due to their red colour, brown dwarfs are a common contaminant for high-redshift sources. The number density of brown dwarfs is, similar to the PISNe discussed above, small. 
    Assuming that there are 25-100 billion brown dwarfs in the Milky Way \citep{muzic17}, and that they follow the ratio of stars in the halo to stars in the whole Galaxy of 10-30\% \citep{pillepich14}, we can estimate their observed number density on the sky: $\Sigma_\mathrm{BD} = f_\mathrm{halo} \times N_\mathrm{BD}/A_\mathrm{sky}\approx 0.2 \times 5\times10^{10}/(4\pi\mbox{\,sr}) \approx 70\,\mathrm{arcmin}^{-2}$, lower than what is predicted by both J19 and V20. We calculate the colours in the NIRCam filters with models by \cite{saumon08}. Only very cloudy brown dwarfs have a higher flux in the F200 than the F444 filter, mildly cloudy or non-cloudy ones are too red to even appear in our colour-colour diagram in Figure \ref{fig:cc}. These very cloudy brown dwarfs show only a small flux difference in the F150 to F200 filter, and hence very few objects will occupy an overlapping region with the \textit{Pop~III + nebula + evolution} model. 
\end{itemize}{}
In summary, we expect the ULT to unambiguously detect 
Pop~III sources. We have investigated other sources, but they are either very rare or emit flux at lower wavelengths than 
minihalos at $z\ge 10$, and can thus be neglected here. 

\subsection{Technical Feasibility}\label{feasibility}

From the previous sections we conclude that an instrument with a field of view of few square arcmin and a sensitivity of 39~mag$_{\mathrm{AB}}$ for point-like sources in the F150 and F200 bands will suffice to detect tens to hundreds of minihalos at redshifts $10-15$. The halos are $< 1$~kpc in physical size, with the light-emitting region likely at least a factor of 10 smaller. Simulations show that the region of active star formation is $<<1$~pc \citep{greif11} while the size of the light-emitting HII region might reach 80~pc maximum size \citep{Anna17}. The angular scale at the redshifts of interest ($z\sim 15)$ is $\sim$3~kpc/arcsec and the diffraction limit for a 100~m aperture is 5~mas at 2~$\mu$m, or 15~pc. While the star-forming regions and the star clusters they contain will be unresolved, the nebular region at peak size might be just resolved by this largest of telescopes given perfect optics. For the purpose of this paper, we can assume that we are dealing with unresolved emission sources.

A telescope capable of reaching these limits has been proposed, for example, by \citet{Angel+2008} in the form of a cryogenic liquid-mirror telescope on the surface of the moon. However, as the authors mention, a liquid mirror might face some challenges, as it is unclear 
what effect lunar dust would have on the instrument and the observations.

To avoid an articulating mount, the telescope would be placed at the lunar pole, constantly pointing at the zenith. To assure thermal isolation, it would be located in the permanent shadow in a lunar crater. The limit on exposure time is then given by the precession of the moon, and is of the order of several days. This can only be extended by the addition of some active tracking facility, for example a moving prime focus platform. However, to reach the sensitivities we require, a mirror diameter of 100\,m is necessary. We note that \citet{Angel+2008} do consider mirror sizes of 20 up to 100\,m in their preliminary design studies. While 100\,m is evidently challenging, it is within the realm of possibility for mid-century technology.

\section{Summary and Conclusions}\label{conclusion}
In this paper, we have explored the requirements for a possible observation of 
Pop~III stars - thus providing a vista into the future beyond JWST. We find that we need to reach an AB~magnitude of $\approx 39{}$ in order to detect these  
first luminous objects in the universe, which would be possible with a 100\,m diameter mirror on the moon \citep{Angel+2008}. The number density of Pop~III sources is very high, so even a small field of view of a few square arcminutes will be sufficient to observe hundreds of minihalos. 
We find that a Pop~III stellar population with nebula emission 
can be observed more easily than a single, massive Pop~III star on the main sequence, and that both show distinct colours and magnitudes from contaminants such as Pop~II stellar populations and brown dwarfs. 

In our analysis, we do not explicitly consider streaming velocities \citep{th10}. 
These relative velocities between gas and dark matter vary over large (Mpc) scales and 
are known to offset the halo mass for star formation to 
larger masses \citep{anna19}. In regions of the universe with a high (low) streaming 
velocity, we expect an under (over) density of star forming minihalos. These spatial variations 
are beyond the scope of this paper. In addition, we here explore two extreme cases for Pop~III stars with maximal and completely absent nebula emission. While the proposed ULT can detect both, more precise statistical predictions of nebula properties will lead to better predictions of the Pop~III observability and the optimal survey strategy, closer to the actual telescope design. A full radiative-transfer model of Pop~III sources is further beyond the scope of this study.

Another possibility to observe fainter objects can be achieved by strong gravitational lensing. Magnifications of factors of tens to hundreds could be reached at high redshift, and a Pop III candidate object has recently been detected \citep{vanzella20}. However, the sky area with these high magnifications is very small, and - if at all - only a few objects would be observable within survey volumes as large as 100 deg$^2$ \citep{zack15}. With the proposed ULT, hundreds of Pop III star forming regions in minihalos are expected to be detected.

\section*{Acknowledgments}
We would like to thank the referee for constructive comments 
and Caroline Morley for input on the brown dwarf spectra. 
Support for this work was provided by NASA through the NASA Hubble Fellowship grant HST-HF2-51418.001-A awarded  by  the  Space  Telescope  Science  Institute,  which  is  operated  by  the Association  of  Universities for  Research  in  Astronomy,  Inc.,  for  NASA,  under  contract NAS5-26555. 
\setlength{\bibhang}{2.0em}
\setlength\labelwidth{0.0em}
\bibliographystyle{mn2e}
\bibliography{refs}
\label{lastpage}

\end{document}